\documentclass[aip,jcp,dvips,floatfix,superscriptaddress,intlimits,twocolumn]{revtex4}

\usepackage{amsmath}
\usepackage{graphicx}
\usepackage{color}

\newcommand{\wn}{\mathrm{cm^{-1}}}
\newcommand{\ps}{\mathrm{ps}}
\newcommand{\ohaae}{j=3/2,F_1,e}
\newcommand{\ohaaf}{j=3/2,F_1,f}
\newcommand{\ohabe}{j=5/2,F_1,e}
\newcommand{\ohabf}{j=5/2,F_1,f}

\newcommand{\ohbaf}{j=1/2,F_2,f}

\begin{document}

\title{Resonances in rotationally inelastic scattering of OH($X^2\Pi$) with helium and neon}
\date{\today}

\author{Koos B. Gubbels}
\email{K.B.Gubbels@science.ru.nl}
\affiliation{Fritz-Haber-Institut der Max-Planck-Gesellschaft, Faradayweg 4-6, D-14195 Berlin,\\ Germany}
\affiliation{
Radboud University Nijmegen, Institute for Molecules and Materials,
Heyendaalseweg 135, 6525 AJ Nijmegen, \\
The Netherlands}

\author{Qianli Ma}
\affiliation{Department of Chemistry, The Johns Hopkins University, Baltimore, Maryland 21218-2685, USA}

\author{Millard H. Alexander}
\affiliation{Department of Chemistry and Biochemistry and
Institute for Physical Science and Technology, University of
Maryland, College Park, MD 20742-2021, USA}

\author{Paul J. Dagdigian}
\affiliation{Department of Chemistry, The Johns Hopkins University, Baltimore, Maryland 21218-2685, USA}

\author{Dick Tanis}
\affiliation{
Radboud University Nijmegen, Institute for Molecules and Materials,
Heyendaalseweg 135, 6525 AJ Nijmegen, \\
The Netherlands}

\author{Gerrit C. Groenenboom}
\affiliation{
Radboud University Nijmegen, Institute for Molecules and Materials,
Heyendaalseweg 135, 6525 AJ Nijmegen, \\
The Netherlands}

\author{Ad van der Avoird}
\affiliation{
Radboud University Nijmegen, Institute for Molecules and Materials,
Heyendaalseweg 135, 6525 AJ Nijmegen, \\
The Netherlands}

\author{Sebastiaan Y. T. van de Meerakker}
\affiliation{
Radboud University Nijmegen, Institute for Molecules and Materials,
Heyendaalseweg 135, 6525 AJ Nijmegen, \\
The Netherlands}

\begin{abstract}
We present detailed calculations on resonances in rotationally
and spin-orbit inelastic
scattering of OH ($X\,^2\Pi, j=3/2, F_1, f$) radicals with He and Ne
atoms. We calculate new \emph{ab initio} potential energy surfaces for
OH-He, and the cross sections derived from these surfaces
compare favorably with the recent crossed beam scattering experiment of
Kirste \emph{et al.} [Phys. Rev. A \textbf{82}, 042717 (2010)]. We
identify both shape and Feshbach resonances in the integral and
differential state-to-state scattering cross sections, and we discuss
the prospects for experimentally observing scattering resonances using
Stark decelerated beams of OH radicals.
\end{abstract}

\maketitle

\section{Introduction}

Measurements of state-to-state cross sections provide important tests of
the reliability of computed potential energy surfaces (PES's) describing
the interaction of atoms and molecules \cite{Bernstein1979}. Cross
sections for collision-induced rotational transitions are sensitive to
the anisotropy of the PES. Since non-bonding interactions are relatively
weak, the magnitudes of the cross sections are mostly sensitive to the
repulsive part of the PES, except at very low collision energies. An
alternative, spectroscopic approach to gaining information on PES's is the
determination of the energies of the bound levels of van der Waals
complexes of the collision partners \cite{Bacic1996, Wormer2000}. The
energies of the bound levels are mainly sensitive to the attractive part
of the PES's. As we go up higher in the manifold of these weakly bound
levels, the energies of these levels eventually become higher than the
dissociation energy of the complex, and such levels are quasi-bound.
These quasi-bound levels are often described as resonances and can be
thought of as a distortion of the continuum in the collision energy
dependence of state-to-state cross sections \cite{Child1974}. In
inelastic scattering, resonances are called shape or orbiting resonances
when the quasi-bound levels involve monomer levels that are the same as in the initial or final level of the collision-induced transition,
or Feshbach resonances when they involve different monomer states
\cite{Bernstein1979,Child1974}. Due to their sensitivity to the PES,
resonances can reveal important information on the PES
\cite{Lique2011,Chandler:JCP132:110901}. So far, however, resonant
structures in scattering cross sections have been experimentally
observed only in exceptional cases
\cite{Schutte:PRL29:979,Schutte:JCP62:600,Toennies:JCP71:614,Qiu:Science311:1440}.

The crossed molecular beam technique has been an extremely useful tool
for the determination of state-to-state cross sections, both integral
and differential, as well as their dependence upon the collision energy
\cite{Levine1987}. The recently developed Stark deceleration technique,
taking advantage of the interaction of polar molecules with time-varying
electric fields, has allowed continuous tuning of the beam velocity
\cite{vandeMeerakker2008}. This has facilitated measurements of the
collision energy dependence of state-to-state integral cross sections
down to energies of $70~\wn$ \cite{Scharfenberg2010}. Moreover, the
velocity spread in such decelerated beams is much smaller than in
conventional molecular beams. Thus far, an energy resolution of
$\ge13~\wn$ has been achieved for collisions of OH radicals with rare
gas atoms \cite{Gilijamse2006, Scharfenberg2010, Kirste2010}. This
resolution is mainly limited by the velocity and angular spread of the
atomic collision partner, and is too low to experimentally resolve
scattering resonances. A recent study has shown that the energy
resolution can be improved significantly by an appropriate choice of the
beam velocities and interaction angle \cite{Scharfenberg2011}. When
these measures are put into practice in the laboratory, collision energy
resolutions can be obtained that may enable the observation of
scattering resonances.

Atom-molecule collisions are the simplest type of collision process in
which rotationally inelastic transitions can be observed. Early
calculations \cite{Erlewein:ZP211:35,Erlewein:ZP218:341} on rotationally
inelastic scattering of N$_2$ molecules with He atoms have shown that
resonances occur at low collision energies, but the experimental
verification of these predictions was not yet possible. Collisions of
OH($X^2\Pi$) with rare gases have emerged as paradigms of scattering of
an open-shell molecule with an atom \cite{Beek2000, Beek2000a, Klos2007,
Pavlovic2009, Dagdigian2009, Dagdigian2009a, Scharfenberg2010,
Kirste2010}. The OH-rare gas systems are good candidates for the
observation and analysis of resonances in rotationally inelastic
collisions because the collision energy can be reduced by Stark
deceleration of the OH beam. Since OH($X^2\Pi$) is an open-shell
molecule with orbital degeneracy, the collision dynamics is governed by
two PES's,\cite{Alexander1985} and interesting multi-state dynamics can
occur. Of particular interest for the study of resonances are the OH-He
and OH-Ne systems, since the dissociation energies of these systems are
smaller than the rotational level spacings of the OH radical. The
resonance features associated with the various rotational levels are
therefore well separated. The shallow van der Waals wells support only
one or two stretch vibrational levels \cite{Lee2000}, resulting in a
rather simple, yet interesting, analysis of the resonances. Shape
resonances in OH-He collisions were previously analyzed by
\citeauthor{Dagdigian2009a}\cite{Dagdigian2009a} in a study of elastic
depolarization. Bound states of the OH-He complex have been
investigated spectroscopically by Han and Heaven, who identified complex
features as scattering resonances in OH(\textit{A})-He
\cite{han:064307}.

Here, we present a detailed and precise study of scattering resonances
in the OH-He and OH-Ne systems, in order to develop insight into the
nature and strength of the resonances and to assist in the experimental
search for such scattering resonances. For the OH-He system, we have
calculated new three dimensional potential energy surfaces. The
collision energy dependence of the relative state-to-state integral
scattering cross sections that are derived from these potentials
compares
more favorably with recent experiments \cite{Kirste2010} than
the results from previous
calculations. For the correct assessment of the resonances, the
calculations are performed on a very fine grid of collision energies,
and particular care is taken to converge the calculations to avoid
numerical artifacts to be interpreted as resonant structures
\cite{footnote1}. We characterize the resonances with various
techniques, including the adiabatic bender model \cite{Holmgren1977,
Alexander1994} and collision lifetime analysis \cite{Smith1960}. We
investigate how the differential cross section for several transitions
changes as the collision energy is scanned through the resonances, and
observe dramatic effects.

This paper is organized as follows: The details of the scattering
calculations are briefly presented in Section~\ref{sec:calc}. In
Section~\ref{subsec:3D-OH-He} we describe the new three-dimensional (3D)
PES's that are developed for OH-He. Section~\ref{sec:heoh_cyb} describes
our calculations on the state-to-state scattering cross sections in
OH-He collisions. A detailed analysis of shape and Feshbach resonances
is given. Section~\ref{sec:neoh} presents similar results for the OH-Ne
system. A discussion of the prospects for observing these resonances in
crossed beam experiments using a Stark decelerator, by either recording
the integral or the differential cross sections, follows in
Sec.~\ref{sec:discussion}.

\begin{figure}
 \centering
 \includegraphics{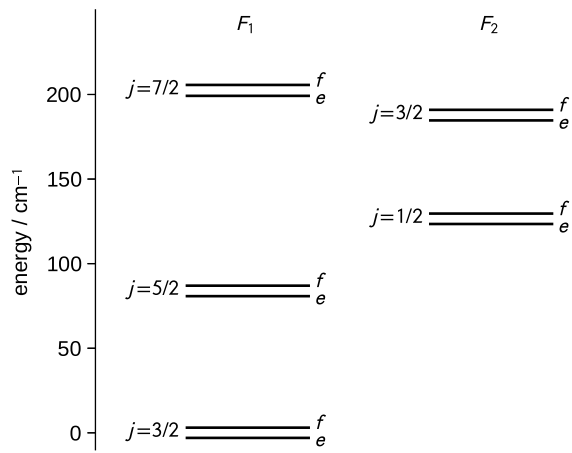}
\caption{Energies of the lower rotational levels of OH($X^2\Pi$). The
$\Lambda$-doublet splitting is exaggerated for clarity. The initial
level for all scattering calculations is the $\ohaaf$ level.}
\label{fig:ohenergies}
\end{figure}

\section{Scattering Calculations}
\label{sec:calc}

The theory of scattering between a molecule in a $^2\Pi$ electronic
state and a structureless atom is well established \cite{Alexander1985}.
The interaction can be described by two PES's corresponding to states of
$A'$ and $A''$ symmetries. For OH-He, we have constructed new PES's,
which are explained in Sect.~\ref{subsec:3D-OH-He}, while for OH-Ne, we
used the PES's by \citeauthor{Sumiyoshi2010} \cite{Sumiyoshi2010}.
Close-coupling calculations were performed both with the \verb:HIBRIDON:
suite of programs \cite{Hibridon}, and with a second independent
scattering program for open-shell diatom-atom scattering described in
Ref. \onlinecite{Scharfenberg2011b}. Care was taken to independently
check the
results with the two scattering programs and to converge the cross
sections. For OH-He the maximum total angular momentum was $J=100.5$ --
$140.5$, depending on the collision energy, and the channel basis
consisted of all rotational levels of OH with $j\le6.5$, while for OH-Ne
the channel basis consisted of all rotational levels with $j\le7.5$. In
this paper, we calculate cross sections from fully converged
close-coupling calculations in order to study resonances in inelastic
collisions between low-lying rotational states of the OH radical. For
reference, the energies of the lower rotational levels of OH($X^2\Pi$)
are displayed graphically in Fig.~\ref{fig:ohenergies}.

\section{3D OH-Helium potential} \label{subsec:3D-OH-He}

A crucial role in the scattering calculations is played by the
interaction potential. In Ref.~\onlinecite{Scharfenberg2011b} a detailed experimental and theoretical
study of inelastic scattering between OH radicals and the rare gas atoms
He, Ne, Ar, Kr and Xe was performed. The theoretical results in that
study were shown to be in excellent agreement with experimentally
measured inelastic cross sections. The agreement between theory and
experiment was, although still very good, the worst for the OH-He
system. It was believed that this was due to the quality of the PES,
since for the OH-He system a smaller basis set was used in the
calculations than for the other systems. For this reason, we construct here a new potential for the OH-He system. We note that in
Ref.~\onlinecite{Scharfenberg2011b} the experimental resolution was
unfortunately not yet high enough to observe resonances.

\begin{figure}
 \centering
 \includegraphics[width=0.6\columnwidth]{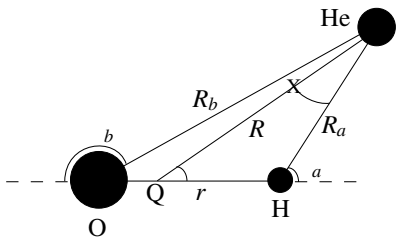}
 \caption{Illustration of the OH molecule and the He atom containing the
 relevant coordinates as used in the fitting of the OH-He potential. $R$
 is the length of the vector ${\bf R}$ that connects the He atom and the
 center-of-mass ($Q$) of the OH molecule, while $\theta$ is the angle
 between ${\bf R}$ and the OH bond vector ${\bf r}$ (pointing from O to
 H) of length $r$. $R_a$ is the length of the vector ${\bf R}_a$ that
 connects the He atom and the H atom, while $\theta_a$ is the angle
 between ${\bf R}_a$ and ${\bf r}$. $R_b$ is the length of the vector
 ${\bf R}_b$ that connects the He atom and the O atom, while $\theta_b$
 is the angle between ${\bf R}_b$ and $-{\bf r}$. The point X marks the
 location of the midbond orbitals.}
\label{figcoor}
\end{figure}

In trying to improve the agreement with the experimental results, we
first constructed new 2D PES's for OH-He collisions. This was done by
enhancing the basis set for the coupled-cluster calculations of the
interaction energy from the augmented triple-zeta correlation-consistent
basis set (AVTZ) used by \citeauthor{Lee2000} \cite{Lee2000} to the
quintuple-zeta basis set (AV5Z). We computed the interaction energies
with the open-shell single and double excitation coupled cluster method
with perturbative triples as implemented in the {\sc molpro} package
\cite{Molpro}. The interaction energies were evaluated for 288
geometries on a two-dimensional grid with 12 Gauss-Legendre points in
the Jacobi-angle $\theta$. The OH bond length was fixed at the
vibrationally averaged distance of $r_0=1.8502$ $a_0$, whereas
\citeauthor{Lee2000} used the equilibrium distance $r_e$. The relevant geometry is illustrated in Fig. \ref{figcoor}. We included
midbond orbitals ($3s$,$3p$,$2d$,$1f$,$1g$) with the exponents of
Ref.~\onlinecite{Koch1998}. These midbond functions were centered on the
vector ${\bf R}$ that connects the He atom and the center-of-mass of the
OH molecule, at a distance from the helium atom that is half the
distance of the helium atom to the nearest atom of the OH molecule. Also the counterpoise
correction of Boys and Bernardi was applied \cite{Boys1970}. The grid of
atom-molecule separations consisted of 18 points ranging from $R=3 \,a_0$
to $9 \,a_0$ at short range and 6 points on an approximately logarithmic
scale up to $25 \,a_0$ at long range.

As mentioned before, two potential energy surfaces
belonging to states of $A'$ and $A''$ symmetry are involved in the
OH-rare-gas atom scattering. The average $V_{\rm s}$ and
half-difference $V_{\rm d}$ of these potentials can be expanded in Racah
normalized spherical harmonics $C_{l,m}$, namely
\begin{eqnarray}\label{eqpot}
\label{eq:potexpand}
V_{\rm s} &=& \frac{V_{A'}+V_{A''}}{2} = \sum^{l_{\rm max}}_{l=0} v_{l,0}(R)C_{l,0}(\theta,0),\nonumber \\
V_{\rm d} &=& \frac{V_{A''}-V_{A'}}{2} = \sum^{l_{\rm max}}_{l=2}
v_{l,2}(R)C_{l,2}(\theta,0),
\end{eqnarray}
where we included all terms up to $l_{\rm max}=11$. In the long range ($R>10 a_0$), the expansion coefficients $v_{l,0}(R)$ were fitted to inverse powers $R^{-n}$ with
$n\ge 6$, namely
\begin{eqnarray}
v^{\rm lr}_{l,0}(R,\theta)= \sum^{11}_{n = n_0(l)} c_{l,n}\frac{f_n(\beta R)}{R^n},
\end{eqnarray}
where we note that the allowed values for $n$
depend on $l$ \cite{Avoird1980}. For example, for $l=0$ we have $n_0(l) = 6$ and only even values of $n$ are present, while for $l=1$ we have $n_0(l) = 7$ and only odd values of $n$ are present. From the fitted coefficients we only kept the leading long-range terms for $l=0$ to $l=4$. We used the Tang-Toennies damping
function \cite{tang:84}
\begin{equation}\label{eqttdf}
f_n(x)=1-e^{-x}\sum_{k=0}^n\frac{x^k}{k!}
\end{equation}
to damp these five long-range terms in the short range with $\beta = 0.6 a_0^{-1}$. In the short range ($R<5.5 a_0$),
the expansion coefficients $v_{l,0}(R)$ were fitted to an exponential, namely
\begin{eqnarray}
v^{\rm sr}_{l}(R)= s_{l} e^{-\alpha_{l} R }.
\end{eqnarray}
The difference between the {\it ab initio} interaction energies and the analytic long range and short range functions was fitted with a
reproducing kernel Hilbert space (RKHS) method \cite{ho:96}. The RKHS
parameter $m$ was chosen such that the RKHS fit would decay faster than
the leading long-range term for each $l$. The RKHS smoothness
parameter was set to $2$. For the expansion coefficients of the difference potential, $v_{l,2}(R)$, no analytic short range and long range fit was performed, so that everywhere the RKHS method was used. Using the described procedure, we obtained an accurate
fit to the {\it ab initio} points. More details of the fit can be found on {\sc epaps} \cite{epaps:pot}, where we provide a {\sc fortran} 77 code for the two-dimensional AV5Z potential.

We found that the absolute minimum of the fitted potential is
located at $\theta =68.7^{\circ}$, $R =5.69 \,a_0$ on the $A'$ PES,
corresponding to an interaction energy of $V_{A'}= -29.8$
cm${}^{-1}$. The minimum potential energy values for $\theta =
0{}^{\circ}$ and $\theta = 180{}^{\circ}$ were found at $R =6.56 \,a_0 $
and $R = 6.09 \,a_0$, giving rise to $V_{A'/A''} = -27.1$ cm${}^{-1}$ and
$V_{A'/A''} = -21.6$ cm${}^{-1}$, respectively. For comparison, we also
mention the values obtained by \citeauthor{Lee2000} \cite{Lee2000}, who
found that the absolute minimum of their potential was located at
$\theta = 68.6 {}^{\circ}$, $R = 5.69 \,a_0$ for $A'$ symmetry, with an
interaction energy of $V_{A'}= -30.0$ cm${}^{-1}$. The minimum values
for $\theta = 0{}^{\circ}$ and $\theta = 180{}^{\circ}$ were found at $R
=6.54 \,a_0 $ and $R = 6.09$, giving rise to $V_{A'/A''} = -27.1$
cm${}^{-1}$ and $V_{A'/A''} = -21.8$ cm${}^{-1}$, respectively. The two
potentials are seen to give very similar results for the local
and global minima. Moreover, using the new AV5Z potential for scattering
calculations, we found only a very slight improvement in the agreement
with the experimental data.

Therefore, we tried to improve the PES further by taking the vibrational
motion of the OH radical into account. To this end, we computed the
interaction energies of the OH-He system on a three-dimensional grid. At
short and intermediate range we used a step size of $\Delta R =
0.25 \,a_0$ for $3 \,a_0 \le R \le 12.5 \,a_0$ and $\Delta r = 0.25
\,a_0$ for $0.75 \,a_0 \le r \le 4.5 \,a_0$. For the angle $\theta$ we used an
equidistant grid of 16 points including $0$ and $180^{\circ}$ with a spacing of
$\Delta \theta = 12^{\circ}$. At long range we used 4 equidistant points
between $14 \le R \le 20$, while we used a step size of $\Delta r = 0.5 \,a_0$
for $0.75 \,a_0 \le r \le 4.25 \,a_0$. The distance $r=4.5 \, a_0$ was also
included in the long-range fit. For the angle $\theta$ we used an
equidistant grid of 9 points with a spacing of $\Delta \theta = 22.5^{\circ}$.
On this grid we computed the interaction energies with a triple-zeta
basis set (AVTZ) and using midbond orbitals with geometry-dependent
exponents \cite{Groenenboom2003}. Especially for large $r$ and small
$R$, the electronic structure calculations did not always
converge. Then, we obtained the energy for the corresponding grid point
by means of interpolation or extrapolation from neighbouring grid
points.

\begin{figure}
\centering
\includegraphics[width=1.0\columnwidth]{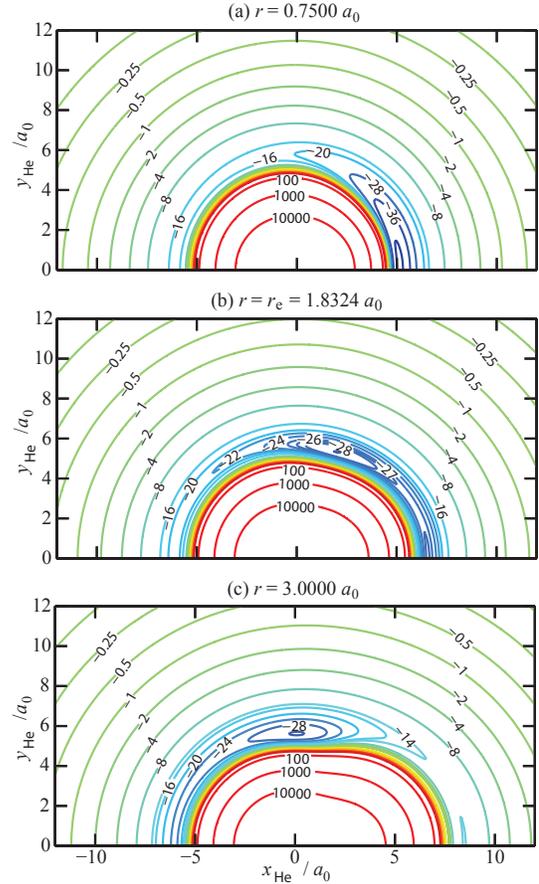}
\caption{$A'$ potential energy surface. The OH radical lies on the
horizontal axis, with the center-of-mass of the molecule at the origin.
The O atom lies left of the origin, the H atom to the right. For each
geometry of the complex, defined by the OH bond length $r$ and the position
$(x_{\rm He},y_{\rm He})$ of the He atom, the interaction energy is
calculated, resulting in contours with the unit of cm${}^{-1}$. The
three plots differ in the OH bond length, namely in panel a) we have
$r=0.75 \,a_0$, in panel b) $r=1.8324 \,a_0$, and in panel c) $r =3.00 \,a_0$.}
\label{figpesap}
\end{figure}

\begin{figure}
 \centering
 \includegraphics[width=1.0\columnwidth]{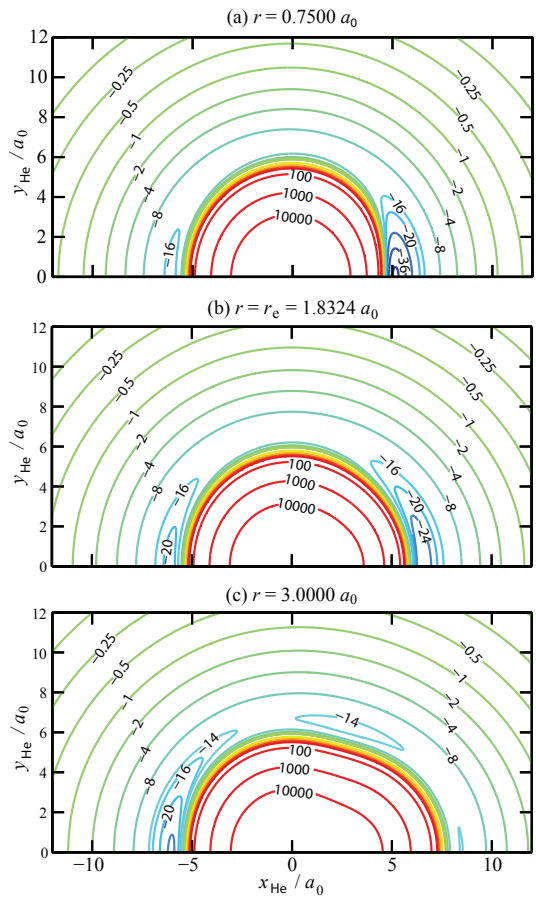}
 \caption{Similar to Fig. \ref{figpesap}, but showing the $A''$ potential energy surface.}
 \label{figpesapp}
\end{figure}

To perform the fit of the sum interaction potential $V_{\rm s}$, we
proceed in the following way. We represent the potential as a sum of
three terms, namely
\begin{eqnarray}\label{eqvs}
&&V_{\rm s}(R,\theta,r)= \nonumber \\
&&V_{\rm s}^{\rm sr}(R_a,\theta_a,r)+V_{\rm s}^{\rm sr}(R_b,\theta_b,r)+V_{\rm s}^{\rm lr}(R,\theta,r),
\end{eqnarray}
where the different coordinates are defined in Fig. \ref{figcoor}. This
representation is convenient because the coordinates of the first and
second term of Eq.~(\ref{eqvs}) are ideally suited to describe the short-range behavior near the H and O atom, respectively, while the
coordinates of the third term are convenient to describe the long-range
behavior. The short-range terms are fitted by
\begin{eqnarray}
&&V_{\rm s}^{\rm sr}(R_i,\theta_i,r)=\sum_{l=0}^{l^0_{\rm max}}e^{-\beta_i R_i}P_{l}( \cos \theta_i) s_l^{(i)}\nonumber\\
&&+\sum_{l=0}^{l^i_{\rm max}}\sum_{k=0}^{k^i_{\rm max}}\sum_{n=0}^{n^i_{\rm max}}R_{i}^n e^{-\beta_i R_i}P_{l}( \cos \theta_i)r^k e^{-\alpha_i r^3} s_{lnk}^{(i)},\quad
\end{eqnarray}
where $i=a,b$, while $P_{l}(x)$ are Legendre polynomials corresponding to
the functions $C_{l,0}(\theta,0)$ of Eq.~\ref{eq:potexpand}. We
used the values $l^0_{\rm max}=1$, $n^a_{\rm max}=3$, $k^a_{\rm max}=8$,
$l^a_{\rm max}=7$, $n^b_{\rm max}=3$, $k^b_{\rm max}=8$ and $l^b_{\rm
max}=5$. The long range term is fitted by
\begin{eqnarray}
V_{\rm s}^{\rm lr}(R,\theta,r)=\sum_{n=6}^{13}\sum_{l=0}^{n-4}\frac{f_n(\beta R)}{R^n}P_{l}( \cos \theta) c_{nl}(r),
\end{eqnarray}
where $f_n$ is the damping function of Eq.~(\ref{eqttdf}). Nonzero
values of $c_{n l}$ occur only for even values of
$l+n$, and then they are given by
\begin{eqnarray}
c_{n l}(r)=c^{0}_{n l}+\sum_{k=0}^{3}r^k e^{-\alpha_n r^3}c_{nlk}.
\end{eqnarray}
We use two different values for $\alpha_n$, namely $\alpha_n =
\alpha_{\rm I}$ for $ 6 \le n \le 9 $, and $\alpha_n = \alpha_{\rm II}$
for $ 10 \le n \le 13 $. For the difference potential $V_{\rm d}$
similar fit functions are used, only now the Legendre polynomials
$P_{l}(x)$ are replaced by associated Legendre functions
$P^2_{l}(x)$ corresponding to the Racah spherical harmonics
$C_{l,2}(\theta,0)$
of Eq.~\ref{eq:potexpand}, so that all sums start with
$l=2$. Moreover, we use
$l^0_{\rm max}=2$, $n^a_{\rm max}=5$, $k^a_{\rm max}=5$, $l^a_{\rm
max}=6$, $n^b_{\rm max}=5$, $k^b_{\rm max}=4$ and $l^b_{\rm max}=6$. The
linear and the nonlinear fit parameters were determined by minimizing a
weighted least-squares error.

By evaluating the analytic representation of the potential on the {\it
ab initio} grid, we were able to compare the fitted energy values with
the {\it ab initio} values. We found that we only obtained a reliable
fit for OH bond lengths $r\le 3\,a_0$. At smaller values of $R$ the
largest
relative error of an analytic value compared to an {\it ab initio} value
for $r\le 3\,a_0$ was $6.67$\% for the sum potential and 1.10\% for the
difference potential. At large $R$, again considering only $r\le
3\,a_0$, the largest relative error was $3.23$ \% for the sum potential
and $3.32$\% for the difference potential. We also calculated the
average relative error, which for the sum potential was $0.32$\%
at short range and $0.49$\% at long range, while for the difference
potential it was $0.04$\% in the short range and $0.90$\% in the long
range. In Fig.~\ref{figpesap}, we show two-dimensional contour plots of
the fitted OH-He $A'$ PES for $r =0.75 \, a_0$, $r =1.8324 \, a_0$ and $r =3.00 \, a_0$,
while in Fig.~\ref{figpesapp} the same plots are shown for the fitted
$A''$ PES. For the equilibrium bond length $r_e =1.8324 \, a_0$, our fit of the
3-dimensional potential is in very close agreement with the PES of
\citeauthor{Lee2000} \cite{Lee2000}, as it should, since both PES's were
calculated with the same {\it ab initio} method using the same
basis set. The absolute potential energy minimum for $r =1.8324 \,a_0$ is
located at $\theta = 69.2{}^{\circ}$, $R = 5.69 \,a_0$ for $A'$
symmetry, leading to an interaction energy of $V_{A'}= -30.0$
cm${}^{-1}$. The minimum values for $\theta = 0{}^{\circ}$ and $\theta =
180{}^{\circ}$ were found at $R =6.55 \,a_0 $ and $R = 6.09$, giving
rise to $V_{A'/A''} = -27.2$ cm${}^{-1}$ and $V_{A'/A''} = -21.7$
cm${}^{-1}$, respectively. A {\sc fortran} 77 code to generate the
interaction potential $V(R,\theta,r)$ is made available as an EPAPS
document \cite{epaps:pot}.

\begin{figure}
 \centering
 \includegraphics[width=1.0\columnwidth]{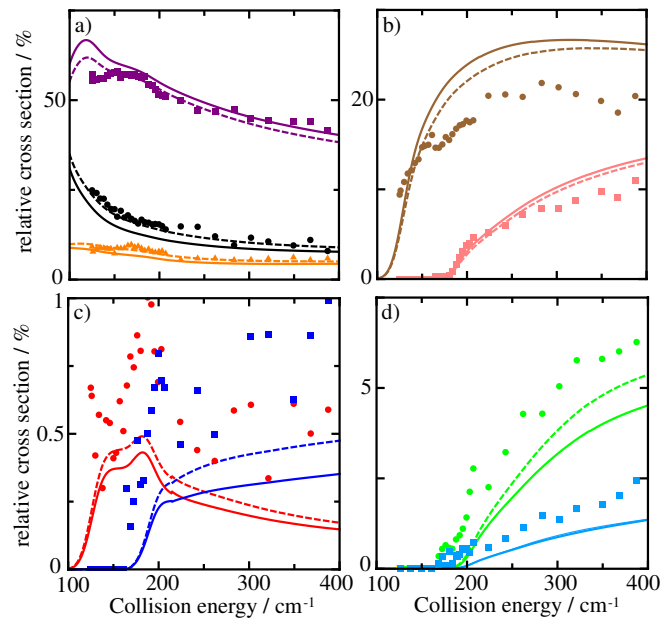}
\caption{Relative state-to-state inelastic scattering cross sections of
OH $(X{}^2 \Pi, j = 3/2, F_1, f)$ radicals with He atoms.
The experimental data points from
Ref.~\onlinecite{Scharfenberg2011b} are shown as dots, while the
theoretically calculated cross sections with the potential of
\citeauthor{Lee2000} \cite{Lee2000} are included as solid curves, and
the results with the adiabatic potential as dashed curves.  On the
vertical axes of the plots, 100~\% refers to the total inelastic cross
section.  Relative cross sections for inelastic collisions populating
the (a) $j=3/2, F_1, e$ (black), $j=5/2, F_1, e$ (purple), and $j=5/2,
F_1, f$ (orange) states; (b) the $j=1/2, F_2, e$ (brown) and $j=3/2,
F_2, f$ (pink) states; (c) the $j=1/2, F_2, f$ (red) and $j=3/2, F_2, e$
(blue) states; (d) the $j=7/2, F_1, e$ (green) and $j=7/2, F_1, f$
(cyan) states.}
\label{fighe}
\end{figure}

To use the three dimensional potential for scattering, we started with
the three-dimensional AVTZ potential, and then subtracted the values of
this potential at $r=r_0$ and added the two-dimensional potential
calculated for $r=r_0$ at the AV5Z level. This implies that the
dependence of the intermolecular potential on the most relevant
coordinates for the scattering calculation, namely $R$ and $\theta$, is
computed at the AV5Z level for $r=r_0$, while the variation of the
potential with the OH bond length $r$ is taken into account at the AVTZ
level. We solved for the OH vibrational motion in the full 3D potential
generated by taking the intermolecular potential energy $V(R,\theta,r)$
and adding the free OH monomer potential $V_{\rm OH}(r)$ \cite{loo:07}.
For fixed $R$ and $\theta$ this leads to an effectively one-dimensional
problem that can be easily solved by standard numerical methods, such as
the discrete variable representation based on sinc-functions (sinc-DVR)
\cite{groenenboom:93}. Taking the resulting ground state energy for each
$R$ and $\theta$ and subtracting the $v=0$ monomer vibrational energy in
the absence of the He atom then results in an adiabatic two-dimensional
PES. We found, actually, that this adiabatic potential is very similar
to the `diabatic' potential obtained by first calculating the lowest
vibrational state of OH in the monomer potential $V_{\rm OH}(r)$ and
then averaging the interaction potential $V(R,\theta,r)$ over this
ground state. The two methods are expected to give similar results since
the vibrational levels of OH are well separated in energy, so that the
weak OH-He interaction gives only a slight admixture of the higher
vibrational states of OH.

The inelastic OH-He cross sections with OH initially in the $\ohaaf$
level were calculated with the adiabatic potential. The results are
shown in Fig.~\ref{fighe}, where the experimental data of
Ref.~\onlinecite{Scharfenberg2011b} is also shown, as well as the
scattering results obtained with the potential of Lee {\it et al.}
\cite{Lee2000}. The theoretical data are convoluted with the
experimental energy resolution. To this end a Gaussian energy
distribution is taken with a standard deviation that is a function of
the energy. The value of the standard deviation ranges from 24
cm${}^{-1}$ at low collision energies to 59 cm${}^{-1}$ at the highest
collision energies. We note that the relative cross sections are
plotted, rather than the absolute cross sections, because these relative
cross sections are experimentally measured. More details can be found in
Refs.~\onlinecite{Scharfenberg2010,Scharfenberg2011b}. We see from Fig.
\ref{fighe} that the overall the agreement with experimental data has
improved noticeably with the adiabatic potential.

\section{OH-HELIUM COLLISIONS}
\label{sec:heoh_cyb}

\subsection{State-to-state integral cross sections}
\label{subsec:ovintc}

\begin{figure}
 \centering
 \includegraphics[width=\columnwidth]{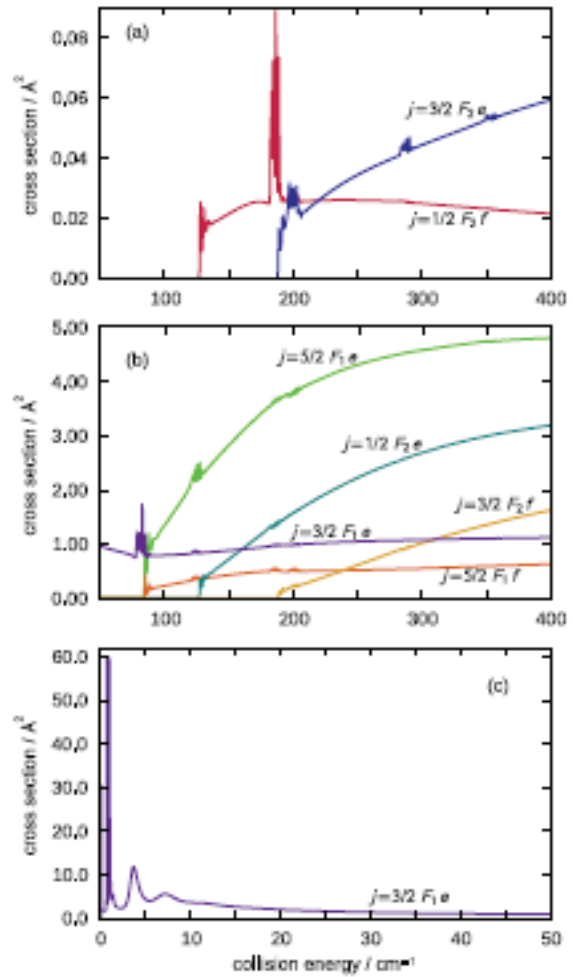}
\caption{State-to-state integral cross sections \emph{vs.}~collision
energy for transitions out of the OH $\ohaaf$ level in collisions with
He. The final levels are indicated for each transition for which the
cross section is plotted.}
\label{fig:ovintc}
\end{figure}

For OH-He collisions, the state-to-state scattering cross sections
were calculated with the adiabatic potential described in the
previous section. In Fig.~\ref{fig:ovintc}, the energy dependence of
the state-to-state integral cross sections for several transitions out
of the $\ohaaf$ level of OH are shown. This level, which is the
higher $\Lambda$-doublet component of the ground rotational level (see
Fig.~\ref{fig:ohenergies}), can be selected with the Stark
decelerator since it is low-field seeking in an inhomogeneous electric
field \cite{Scharfenberg2010}. The cross sections are computed on a very
fine grid of energies to be able to study resonant features in detail.

Away from the resonances, these results are in good agreement with those
previously reported by K{\l}os \emph{et al.} \cite{Klos2007}. As noted
by these authors, there is a propensity for transitions preserving the
total parity. The cross sections are found to be smaller for transitions
with large energy gaps. The initial and final levels of the two
transitions shown in Fig.~\ref{fig:ovintc}(a) have a rather large
energy separation ($>100~\wn$), and the total parity is inverted during
the transitions. Hence, the cross sections for these transitions are
small.

Resonances can be observed in Fig.~\ref{fig:ovintc} near the collision
energies corresponding to thresholds for excitation of the OH radical to
higher rotational and spin-orbit levels. Both shape resonances, which
appear right above the threshold energies for the final levels, and
Feshbach resonances, which appear near the energies where higher
rotational levels than the considered outgoing channel become open, are
present. Except for the $\ohaaf\rightarrow\ohaae$ and the
$\ohaaf\rightarrow\ohbaf$ transitions, the Feshbach resonances are not
significant compared with the background continuum. The
$\ohaaf\rightarrow\ohaae$ transition dominates at low collision energies
and also gives rise to shape resonances with cross sections peaking
above $10~\mathrm{\mathring{A}^2}$. However, these shape resonances
occur at collision energies of only a few wavenumbers.

In the following subsections, we analyze the shape resonances in the
$\ohaaf\rightarrow\ohabe$ transition and the Feshbach resonances in the
$\ohaaf\rightarrow\ohbaf$ transition. The former transition has a large
cross section; the latter transition exhibits strong resonances that
show the largest enhancement compared to the background.

\subsection{Shape resonances}
\label{subsec:shape}

\begin{figure}
 \centering
 \includegraphics{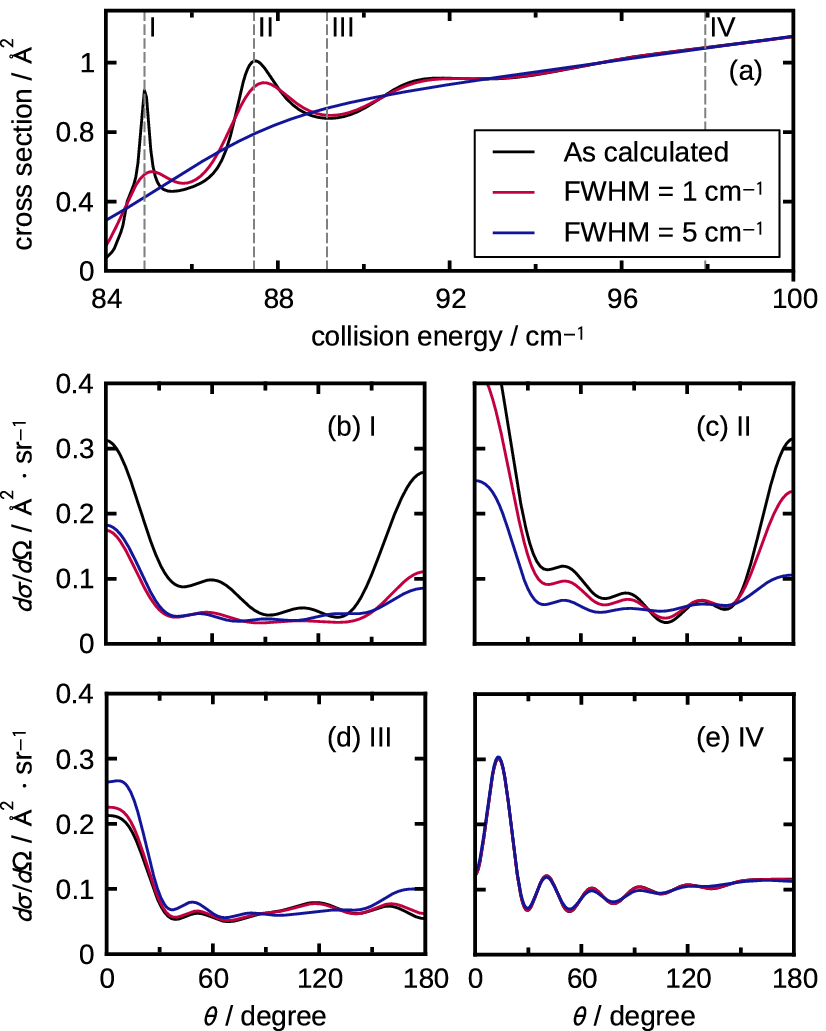}
\caption{(a) State-to-state integral cross sections \emph{vs.}~collision
energy for the $\ohaaf\rightarrow\ohabe$ transition of OH in collisions
with He (thick line), and the integral cross section convoluted with
Gaussian energy distributions of FWHM of $1$ and $5~\wn$ (thin lines).
(b)--(e) Differential cross sections $d\sigma/d\Omega$ of the above
transition at several energies marked as dashed lines with Roman
numerals in (a), together with the differential cross sections
convoluted with Gaussian energy distributions as in (a).}
\label{fig:shape}
\end{figure}

Shape resonances result from quasi-bound states of the van der Waals
complex formed by the collision partners at energies just above the
threshold for the final level. All integral cross sections plotted
in Fig.~\ref{fig:ovintc} display shape resonances. In this subsection we
analyze the shape resonances associated with the
$\ohaaf\rightarrow\ohabe$ transition since it has a large integral cross
section. Figure~\ref{fig:shape}(a) displays these resonances on an
expanded energy scale. Several maxima, with increasing peak width
\emph{vs.}~energy, can be observed, as was also previously found
\cite{Dagdigian2009a} for OH-He and other He-molecule systems
\cite{Dagdigian2010, Lique2011, Dagdigian2011_ch3he}.

To gain more insight, we employ the adiabatic bender model
\cite{Holmgren1977, Alexander1994} to analyze the shape resonances. The
method is similar to the previous analysis of OH-He collisions by
\citeauthor{Dagdigian2009a} \cite{Dagdigian2009a}, except that we used a
close-coupling channel basis instead of a coupled-states one. The full
Hamiltonian with the inclusion of Coriolis coupling and only the
radial nuclear kinetic energy excluded is diagonalized as a function of
$R$. The eigenvalues define a set of adiabatic bender potential energy
curves, which are labeled by the total angular momentum $J$, the total
parity $p$ of the scattering wavefunction, and the cardinal index $n$ of
the eigenvalue. In this paper, we will use the symbol $J^{(+)}_n$ and
$J^{(-)}_n$ to label close-coupling adiabatic bender curves with $p=+1$
and $p=-1$, respectively.

\begin{figure}
 \centering
 \includegraphics{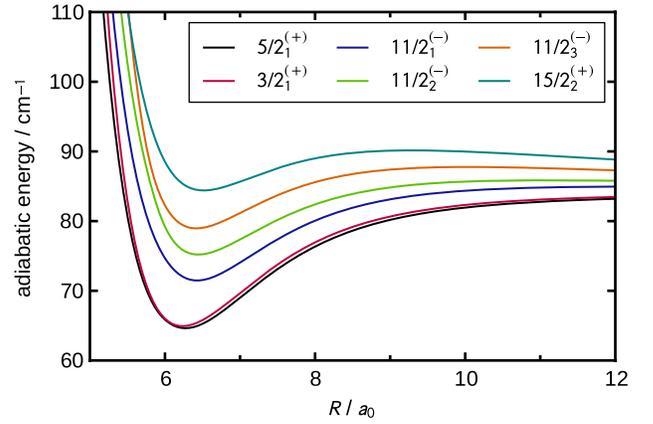}
\caption{Plots of the OH-He adiabatic bender curves that correlate with
the OH $\ohabe$ level, obtained from close-coupling calculations. Curves
are labeled with $J^{(p)}_n$, where $J$, $p$, and $n$ are the total
angular momentum, the total parity of the scattering wavefunction, and
the cardinal index, respectively.}
\label{fig:adiab_shape}
\end{figure}

Figure~\ref{fig:adiab_shape} shows several adiabatic bender curves that
correlate with the OH $\ohabe$ level. The curves marked with
${5/2}^{(+)}_1$
and ${3/2}^{(+)}_1$ are the two lowest lying adiabatic bender curves,
each of which supports only one bound stretch level, with energies of
$77.47~\wn$ and $78.15~\wn$, respectively. As $J$ and $n$ increase, the
curves move up in energy and the well depths become smaller. As a
consequence, some of the bound levels become quasi-bound, and for the
high lying curves (for example, the ${15/2}^{(+)}_2$ curve shown in
Fig.~\ref{fig:adiab_shape}) the wells are too shallow to support any
quasi-bound levels.

\begin{figure*}
 \centering
 \includegraphics{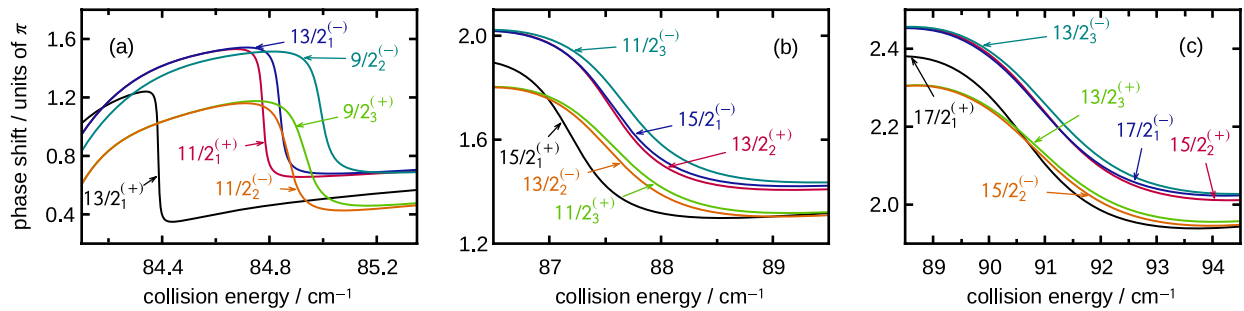}
\caption{Phase shifts as a function of collision energy for
OH($\ohabe$)-He collisions, obtained from close-coupling adiabatic
bender curves described in the text and shown in
Fig.~\ref{fig:adiab_shape}. Curves are labeled with $J^{(p)}_n$, where
$J$, $p$, and $n$ are the total angular momentum, the total parity of
the scattering wavefunction, and the cardinal index, respectively.}
\label{fig:phase_shape}
\end{figure*}

To compute the energies of the shape resonances, we treat the adiabatic
bender curves in conventional one-dimensional scattering problems and
calculate the phase shift. We should be able to observe rapid changes by
$\pi$, signifying resonances, in the collision energy dependence of
phase shift \cite{Child1974}. Fig.~\ref{fig:phase_shape} shows the
phase shift as a function of collision energy for all the adiabatic
bender curves that have such a feature. We see from
Fig.~\ref{fig:phase_shape} that resonances in six adiabatic bender
curves contribute to each of the three peaks shown in
Fig.~\ref{fig:shape}(a) (labeled as I, II, and III), which occur at
$84.8$, $87.6$ and $91~\wn$, respectively. The resonance features in
Fig.~\ref{fig:shape}(a) and Fig.~\ref{fig:phase_shape} match well both
in energy and width. Note that in order to distinguish the phase shift
in different adiabatic bender curves, Fig.~\ref{fig:phase_shape}(b) and
(c) do not show the whole range of the resonances, and thus the resonant
changes in phase shift shown are less than $\pi$.

It is also interesting to compare the differential cross sections for
collision energies on and off a resonance.
Figures~\ref{fig:shape}(b)--(e) display differential cross section for
several energies marked in Fig.~\ref{fig:shape}(a) with Roman numerals.
The center of the two major peaks are marked as I and II, while III and
IV correspond to non-resonant energies. We observe significant backward
scattering for energies I and II, likely because of an increased time
delay of collision due to the formation and decay of quasi-bound levels
of the van der Waals complex. Backward peaks are insignificant for
energies III and IV. We will further discuss this topic in the next subsection.

\subsection{Feshbach resonances}
\label{subsec:feshb}

\begin{figure}
 \centering
\includegraphics[width=\columnwidth]{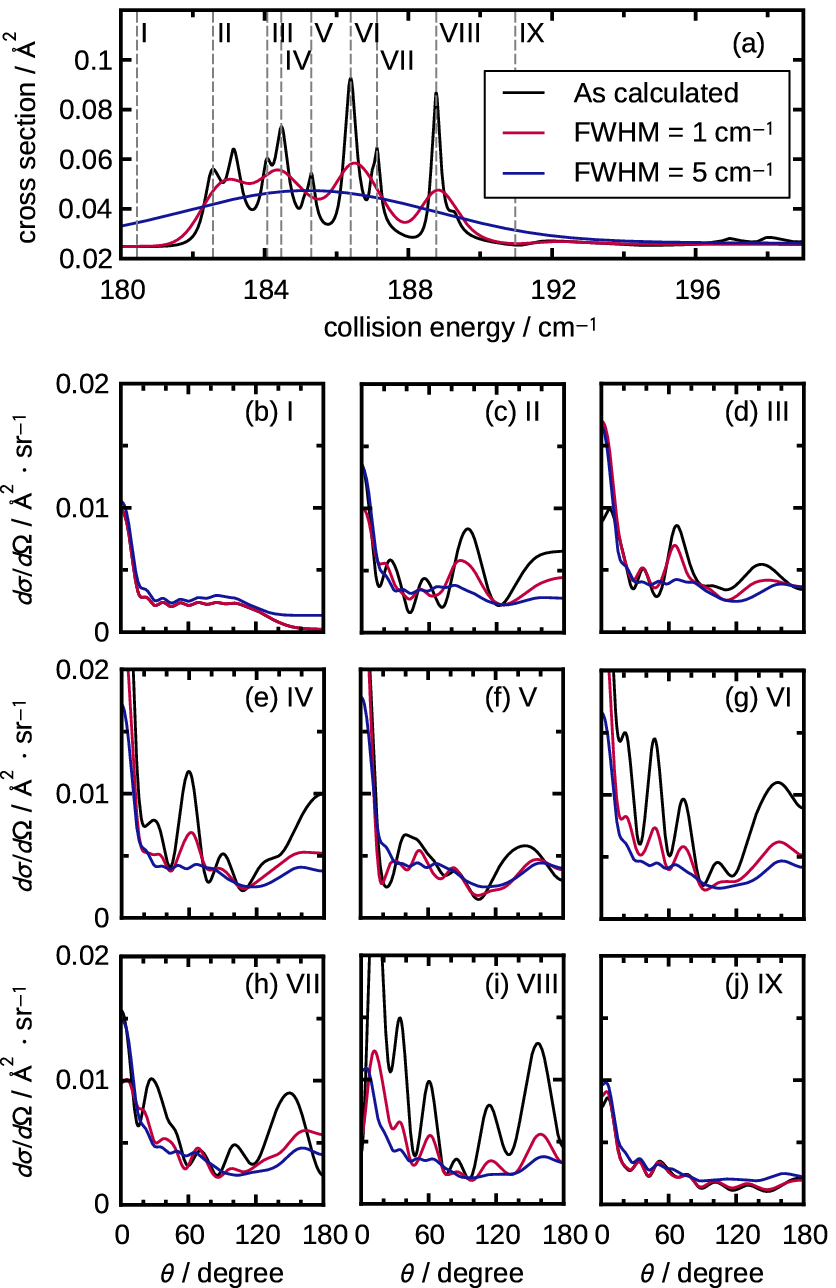}
\caption{(a) State-to-state integral cross section \emph{vs.}~collision
energy for the $\ohaaf\rightarrow\ohbaf$ transition of OH in collisions
with He (black line), and the integral cross section convoluted with
Gaussian energy distributions with FWHM of $1$ (red line) and $5~\wn$ (blue line).
(b)--(j) Differential cross sections $d\sigma/d\Omega$ of the above
transition at several energies marked as dashed lines and Roman numerals
in (a), together with the convoluted differential cross sections with
Gaussian energy distribution as in (a).}
\label{fig:feshb}
\end{figure}

In Feshbach resonances, quasi-bound levels of the collision complex
associated with a given rotational level dissociate to yield the
molecule in a lower-energy rotational level. We consider here Feshbach
resonances associated with the $\ohaaf\rightarrow\ohbaf$ transition.
This transition was chosen for detailed study since the resonance
features show a \emph{ca.}~4-fold increase over the continuum background
(see Fig.~\ref{fig:ovintc}). Figure~\ref{fig:feshb}(a) displays these
resonances on an expanded energy scale.

\begin{figure}
 \centering
 \includegraphics{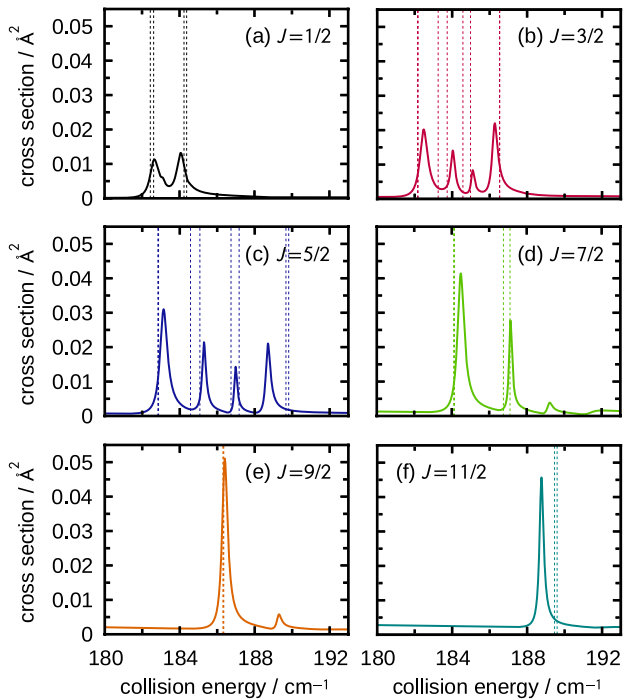}
\caption{Partial cross sections \emph{vs.}~collision energy for the
$\ohaaf\rightarrow\ohbaf$ transition of OH in collisions with He for
total angular momentum $J\le11/2$. The dotted vertical lines denote the
computed energies of the van der Waals stretch levels supported by the
close-coupling adiabatic bender curves.}
\label{fig:partc_feshb}
\end{figure}

It is seen that a rich set of Feshbach resonances exists in a collision
energy range of several $\wn$ below the energetic threshold for opening
of the $F_2, j=3/2$ level at 188 $\wn$. Figure~\ref{fig:partc_feshb}
displays the contribution to the integral cross section for the
$\ohaaf\rightarrow\ohbaf$ transition from various values of the total
angular momentum $J$ (partial cross sections). The individual partial
cross sections exhibit one or more peaks, and their energies shift
toward higher collision energy as $J$ increases. For $J\ge13/2$, no
significant resonances can be found in the energy dependence of partial
cross sections.

We performed an adiabatic bender analysis similar to that described in
subsection \ref{subsec:shape}. We calculated adiabatic bender
potentials by diagonalizing the Hamiltonian expressed in a
close-coupling channel basis. Since all possible values of $l$ (the
orbital angular momentum of the van der Waals complex) are included in
the channel basis, there are multiple adiabatic bender curves for each
value of $J$. These adiabatic bender curves look similar to those shown
in Fig.~\ref{fig:adiab_shape} and are not plotted here. The energies of
the van der Waals stretch levels supported by these curves were derived
using a fixed step-size discrete variable representation (DVR)
method\cite{Colbert1992, Tao2007}. To treat levels that might be
quasi-bound, an infinite barrier was placed at the maximum of the
centrifugal barrier on each adiabatic bender potential. For curves
associated with large $J$, this approximation will lead to calculated
energies higher than they should be and could lead to significant error.
These computed energies are shown as dotted lines in
Fig.~\ref{fig:partc_feshb}. There is a reasonable match between the
energies of the resonances and of the bend-stretch levels, especially
for small $J$.

Fig.~\ref{fig:feshb}(b)--(j) display the differential cross section
for the OH $\ohaaf\rightarrow\ohbaf$ transition at several energies
marked on Fig.~\ref{fig:feshb}(a) with Roman numerals. The energies at I
and IX are not at a resonance, and the differential cross sections show
little backward scattering, while II -- IV correspond to resonance
energies, for which some backward scattering can be observed. The shapes
of the differential cross sections are quite different at resonance
energies compared to collision energies away from the resonances.

\begin{figure}
 \centering
\includegraphics{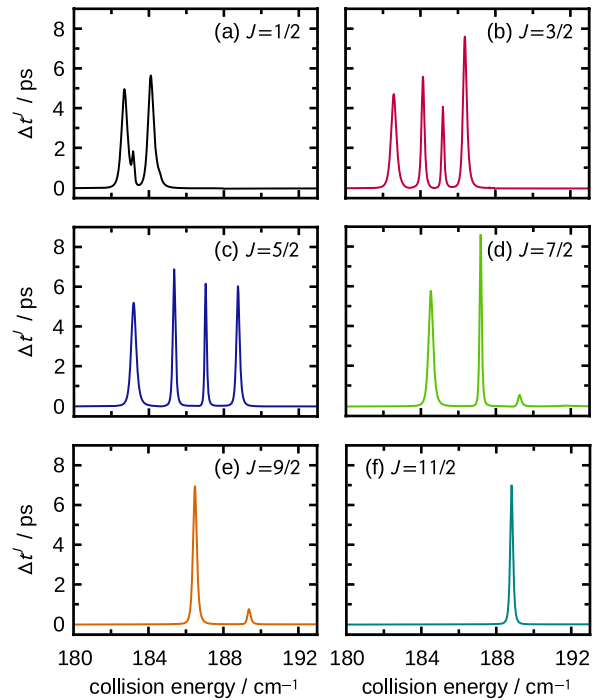}
\caption{Collision lifetime $\Delta t^J(E)$ as a function of collision
energy for the OH $\ohaaf\rightarrow\ohbaf$ transition in collisions
with He, as defined in Eq. (\ref{eq:lifetime}) of the text, for total
angular momenta $J$ = $1/2$ -- $11/2$.}
\label{fig:lifetime_feshb}
\end{figure}

A simple way to analyze the resonances and to qualitatively explain
backward scattering appearing in differential cross sections is to
calculate the collision lifetime, which is the difference between the
time that the collision partners spend in each other's neighborhood with
and without the interaction \cite{Wigner1955, Smith1960, Castillo1996}.
For a direct comparision with the partial cross sections shown in Fig.
\ref{fig:partc_feshb}, we compute the collision lifetime from initial
state $\gamma$ to final state $\gamma'$ for individual total angular
momenta $J$, defined as
\begin{equation}
\label{eq:lifetime}
 \Delta t_{\gamma\gamma'}^J(E) = \text{Re} \left[ -i\hbar \sum_{l, p, l', p'} \delta_{pp'} \left( S_{\gamma,\gamma',l,l'}^{J} \right)^{\ast} \frac{dS_{\gamma,\gamma',l,l'}^J}{dE} \right]
\end{equation}
where $l$, $p$ and $l'$, $p'$ denotes the orbital angular momentum and
parity of initial and final levels, respectively, and
$S_{\gamma,\gamma',l,l'}^J$ denotes $S$-matrix elements for total
angular momentum $J$ from close-coupling calculations. The lifetimes
\emph{vs.}~$J$ for the $\ohaaf\rightarrow\ohbaf$ transition are plotted
in Fig.~\ref{fig:lifetime_feshb}. Clearly, the resonance peaks in
Fig.~\ref{fig:partc_feshb} are well reproduced, with collision lifetimes
of a few picoseconds. The most intense resonance peak lies at
$186.4~\wn$, which was largely due to the $J=9/2$ partial cross section.
From Fig.~\ref{fig:lifetime_feshb} we see that the corresponding
lifetime is about 6 ps. We can compare this collision lifetime with the
rotational period of the OH-He van der Waals complex. We estimate the
rotational constant of the complex to be $6.4\times10^{-24}~\mathrm{J}$
from the expectation value of $1/R^{2}$ computed with the wave function
obtained from the DVR method on the lowest lying $J=9/2$ adiabatic
bender curve. This corresponds to a rotational period $14.9~\ps$,
assuming $l=3$.

We thus conclude that the collision lifetime has the same order of
magnitude as the rotational period of the OH-He complex. It is
therefore not surprising to observe significant backward scattering at
some resonance energies. At off-resonance energies the collision
lifetime will be $\ll1~\ps$, which is much smaller than the OH-He
rotational period. Hence, backward scattering is expected to be barely
observable.

\section{OH-NEON COLLISIONS}
\label{sec:neoh}

\begin{figure}[t]
 \centering
 \includegraphics[width=0.8\columnwidth]{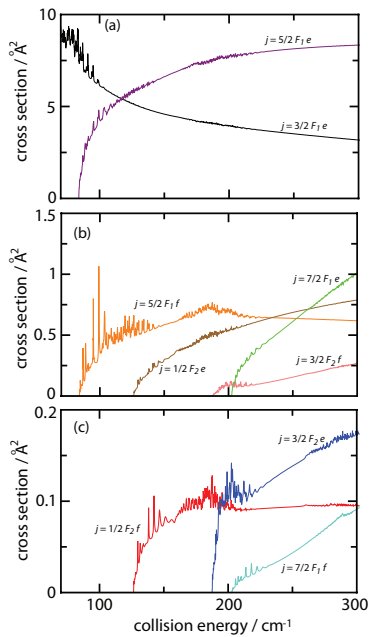}
\caption{State-to-state inelastic scattering cross sections of OH
$(X{}^2 \Pi_{3/2}, j = 3/2, f)$ radicals with Ne atoms as a function of
the collision energy. Cross sections for inelastic collisions populating
the (a) $j=3/2, F_1, e$ (black) and the $j=5/2, F_1, e$ (purple) states;
(b) the $j=5/2, F_1, f$ (orange), $j=1/2, F_2, e$ (brown), $j=3/2, F_2,
f$ (pink), and the $j=7/2, F_1, e$ (green) states; (c) the $j=1/2, F_2,
f$ (red), $j=3/2, F_2, e$ (blue), and the $j=7/2, F_1, f$ (cyan)
states.}
\label{figne}
\end{figure}

\begin{figure*}[t]
\centering
\includegraphics[width=1.5\columnwidth]{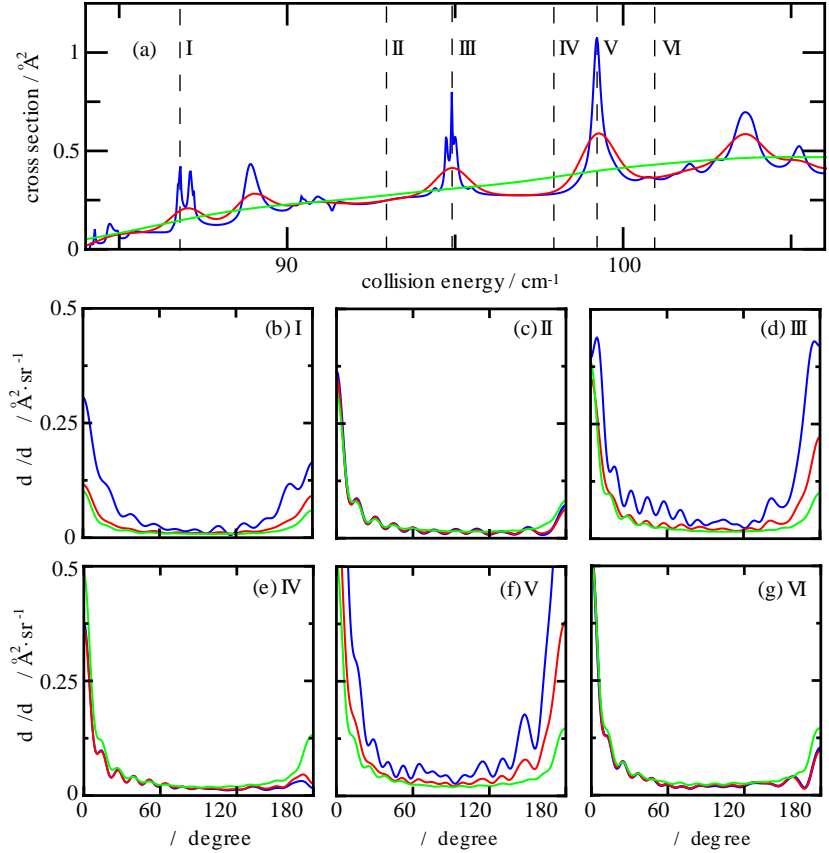}
\caption{(a) Integral cross section for collisions of OH radicals with Ne
atoms as a function of the collision energy. The initial state of the OH
radical is the $j = 3/2, F_1, f$ state and the final state is the
$j=5/2, F_1, f$ state. The blue curve shows the theoretical results
without any convolution, the red curve shows the integral cross section
convoluted with a Gaussian energy distribution having a FWHM of 1
cm${}^{-1}$, while for the green curve a FWHM of 5 cm${}^{-1}$ was used.
At the collision energies designated with the Roman numerals I-VI,
differential cross sections were calculated, shown in panels (b) to (g).
Again the blue curves are not convoluted, while for the red curves a
FWHM of 1 cm${}^{-1}$ was used, and for the green curves a FWHM of 5
cm${}^{-1}$ At the scattering resonances (I, III and V), strong
backscattering is observed.}
\label{figneres}
\end{figure*}

To describe the interaction between OH and Ne, we used the PES
of \citeauthor{Sumiyoshi2010} \cite{Sumiyoshi2010}. This PES was
calculated using an explicitly correlated, spin-unrestricted
coupled-cluster approach [UCCSD(T)-F12b] with a quintuple-zeta basis set
(AV5Z). Although \citeauthor{Sumiyoshi2010} calculated a
three-dimensional potential, we used in Ref.~\cite{Scharfenberg2011b}
their interaction potential evaluated at the equilibrium distance
$r_e = 1.832 \,a_0$ for the scattering calculations, so that no effect of the OH vibrational motion was included. In that reference it was shown that this procedure already gives excellent agreement between theory and high-precision scattering experiments for OH-Ne collisions. Since we found in Section \ref{subsec:3D-OH-He} that the vibrational motion of OH can be of quantitative influence, we also calculated an adiabatic potential from the three-dimensional potential of \citeauthor{Sumiyoshi2010} in the same way as we did for OH-He. The resulting adiabatic potential was found to improve slightly the
excellent agreement with the experimental results for the scattering of OH and Ne. In the present study, we use the adiabatic potential throughout and compute the cross sections on a much finer grid than in the study by
\citeauthor{Scharfenberg2011b} \cite{Scharfenberg2011b} in order to
study scattering resonances. In Fig. \ref{figne} we show the energy
dependence of state-to-state integral cross sections for collisions of
the OH radical with Ne atoms, where the OH radicals are initially in the
$\ohaaf$ level. Overall, the behavior of the inelastic cross sections as
a function of energy is rather similar to what was observed for the
OH-He system in Section \ref{sec:heoh_cyb}. For example, we again
observe a propensity for transitions preserving the total parity.
However, in the OH-Ne system none of the channels appears to have
particularly strong Feshbach resonances, as was the case for the
$\ohaaf\rightarrow\ohbaf$ transition of the OH-He system. The most
pronounced resonant features observed for OH-Ne collisions are shape
resonances in the $\ohaaf\rightarrow\ohabf$ transition. In
Fig.~\ref{figneres} we show these shape resonances in more detail.

Looking closely at Fig.~\ref{figneres}, we see several resonance peaks
that correspond to an increase in the cross section by about a factor of
two compared to the nonresonant energies. A relatively strong resonance
occurs at a collision energy of 99.23 cm${}^{-1}$; this resonance
increases the cross section by a factor of four compared to the
background. The latter resonance is indicated by the Roman numeral V.
The main contributions to this resonance originate from partial cross
sections with total angular momenta of $J = 37/2$ and $J = 39/2$. In
Fig.~\ref{figneres}, we also show the differential cross sections for
several energies that are marked by Roman numerals in panel (a). For the
resonances at collision energies of 86.83, 94.90 and 99.23 cm${}^{-1}$,
the cross sections are shown in the panels (b), (d) and (f). In these
plots, large amplitudes for backscattering are observed. To compare, the
differential cross sections were also calculated away from the
resonances at the energies 93.00, 98.00 and 101.00 cm${}^{-1}$, and the
results are shown in the panels (c), (e) and (g). In the case of
nonresonant scattering, the observed backscattering is significantly
reduced. The differential cross sections at these resonances look
similar to those at the shape resonances for the
$\ohaaf\rightarrow\ohabe$ transition of the OH-He system (see Fig.
\ref{fig:shape}), where also an increase in backscattering was found.
With a measurement of the differential cross sections, the strong
increase and decrease in the backscattering might help in experimentally
identifying the shape resonances at 94.90 and 99.23 cm${}^{-1}$.
However, we note that one must be careful with identifying the
backscattering signal with resonances in the cross section. Namely, in
Fig. \ref{figneres}(a) also less pronounced resonances are seen, and not
all of them have such a strong backscattering signal as the strongest
resonances I, III and V. Moreover, closer to the threshold of the
$\ohaaf\rightarrow\ohabe$ transition significant backscattering is
observed away from the resonances.

\section{DISCUSSION}
\label{sec:discussion}
The experimental observation of resonance structures as discussed in
this paper would comprise a very detailed test for the calculated PES's
and scattering calculations on these PES's. The Stark deceleration
technique provides a source of state-selected molecules with a tunable
velocity and narrow velocity distribution. This technique enables
state-to-state scattering experiments in which the collision energy can
be precisely tuned over a wide range with a high collision energy
resolution. Yet, the observation of scattering resonances requires an
energy resolution that has not yet been achieved in this type of
experiments. In this section, we analyze the collision energy
resolution required to observe resonance features in either the
state-to-state integral or the differential cross sections for OH-He and
OH-Ne collisions. We discuss the requirements on the beam velocity
and angular distributions, and discuss the feasibility of obtaining
these distributions.

\begin{figure}[t]
\centering
\includegraphics[width=\columnwidth]{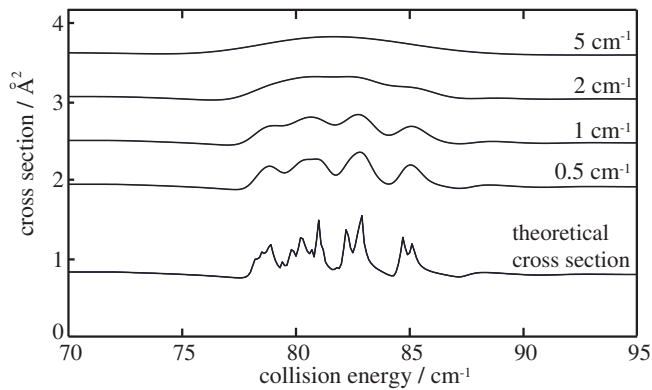}
\caption{Inelastic scattering cross section for the $j=3/2, F_1, f
\rightarrow j=3/2, F_1,e$ channel in collisions of OH radicals with He
atoms, showing Feshbach resonances that correspond to the opening of the
$j=5/2, F_1$ rotational levels of OH. The theoretical curve is
convoluted with Gaussian energy distributions with FWHM of
0.5~cm$^{-1}$, 1~cm$^{-1}$, 2~cm$^{-1}$, and 5~cm$^{-1}$.
The upper four curves have been given a vertical offset for reasons of clarity.}
\label{fig:resolution}
\end{figure}

Referring back to Figs. \ref{fig:shape}, \ref{fig:feshb}, and
\ref{figneres}, the most prominent resonance structures are found for
OH-He in the $j=3/2, F_1, f \rightarrow$ $j=3/2, F_1, e$, $j=3/2, F_1, f
\rightarrow$ $j=5/2, F_1, e$ and the $j=3/2, F_1, f \rightarrow$ $j=5/2,
F_2, f$ transitions. For OH-Ne collisions the $j=3/2, F_1, f
\rightarrow$ $j=5/2, F_1, f$ channel is of most relevance. To simulate
what would be observed in a molecular beam scattering experiment, the
integral cross sections in Figs. \ref{fig:shape}, \ref{fig:feshb}, and
\ref{figneres} are convoluted with Gaussian collision energy
distributions of $1$ and $5~\wn$ full width at half maximum (FWHM). In
Fig. \ref{fig:resolution}, the resonance structure at collision
energies around 85 cm$^{-1}$ for the $j=3/2, F_1, f \rightarrow$ $j=3/2,
F_1, e$ channel in OH-He collisions is shown. This scattering channel
displays a number of Feshbach resonances, corresponding to the opening
of the $j=5/2, F_1$ channels, that are grouped within a relatively
narrow range of collision energies. The theoretical curve is convoluted
using $0.5$, $1$, $2$, and $5$ cm$^{-1}$ (FWHM) energy distributions. We
will use this scattering channel as a benchmark to establish the
energy resolution required in the experiments to observe
signatures of scattering resonances. Scattering resonances are
partially resolved for energy resolutions of $\leq$ 1 cm$^{-1}$. When
a resolution between 1 and 2 cm$^{-1}$ is achieved, some of the
resonance structure is resolved, while for resolutions of $5~\wn$, most
of the resonance structure has disappeared. In these cases, at best only
a broad peak is observed in the integral cross section.

For given beam velocity and angular spread, the collision energy
distribution is a function of the collision energy; the highest
resolutions are obtained for the lowest collision energies. We can
estimate the beam parameters required to reach collision
energy distributions of $\leq$ 2~cm$^{-1}$ at a collision energy of
85~cm$^{-1}$ for collisions of Stark-decelerated OH radicals with He
atoms, i.e., to (partially) resolve the scattering resonances shown
in Fig. \ref{fig:resolution}. We assume a beam intersection angle
of 45$^{\circ}$ and choose the velocities of the He and OH
beams such that the relative velocity vector is perpendicular to the He
atom velocity vector. In this geometry, the collision energy
distribution is almost independent of the He atom beam velocity spread
\cite{Scharfenberg2011}. A collision energy of 85~cm$^{-1}$ is reached
for He and OH velocities of 790~m/s and 1120~m/s, respectively. In the
chosen geometry, the most critical parameter that determines the
collision energy resolution is the distribution in beam intersection
angles $\Delta \phi$. If we assume extremely well collimated beams such
that $\Delta \phi=$ 10~mrad (corresponding to $0.6^{\circ}$), a
velocity spread of the OH radicals of 5~m/s results in a collision
energy resolution of 1.9~cm$^{-1}$.

Experimentally, the most challenging requirement is the angular spread
of both beams. Multiple collimation slits for both the OH and He beams
are required to reach the required angular spreads. The required He atom
velocity can be obtained using a cryogenic source that is maintained at
a temperature of 60 K, and the required OH velocity can be produced
using the Stark decelerator. The required velocity spread for the OH
radicals can be obtained using the Stark decelerator, either by
choosing the appropriate phase angle in the decelerator
\cite{Meerakker:NatPhys4:595}, or by additional phase-space manipulation
techniques \cite{Crompvoets:PRL89:093004}. For the OH-Ne system, even
more stringent requirements apply to the beam distributions due to the
higher reduced mass for this system. Beam speeds of 664~m/s and 470~m/s
for the OH radical beam and Ne atom, respectively, will result in a
collision energy of 85~cm$^{-1}$ using a beam crossing angle of
45$^{\circ}$. For the velocity and angular distributions used above for
OH-He, a collision energy distribution of 2.1 cm$^{-1}$ is obtained.

The signatures of scattering resonances can also be inferred from
differential cross sections. State-to-state differential cross sections
can be measured using the velocity map imaging technique, that provides
the full angular and velocity distribution of the scattered molecules
\cite{footnote2}. Alternatively, information on the differential cross
section may be obtained via Doppler profile measurements of the scattered molecules. The backward scattered components that appear in the
differential cross sections when a resonance is accessed offers
interesting prospects to reveal the existence of resonances. As the
collision energy is tuned over a group of scattering resonances, the
presence of these resonances can in principle be inferred from the
measured product flux at backward scattering angles. In Figs.
\ref{fig:shape}, \ref{fig:feshb}, and \ref{figneres}, the differential
cross sections are shown at collision energies near and at the
resonances, convoluted with Gaussian collision energy distributions of
$1$ and $5~\wn$ full width at half maximum (FWHM). In particular for the
$j=3/2, F_1, f \rightarrow j=5/2, F_1, f$ channel in OH-Ne, a
significant scattering intensity at backward scattering angles remains,
even for a collision energy resolution as high as 5 cm$^{-1}$. In the
integral cross section, no signature of the scattering resonances is
observable at these energy resolutions. In these cases, it may be
favorable to experimentally explore the existence of resonances via
measurements of differential cross sections instead of integral cross
sections.

The examples treated above show that collision energy resolutions in the
1 -- 2 cm$^{-1}$ range, although very challenging experimentally, should
allow for the observation of both shape and Feshbach resonances in the
integral cross sections for inelastic collisions between OH radicals and
He or Ne atoms. The OH radical is an excellent candidate in these
experiments, as a sensitive detection scheme, appropriate for ion imaging,
has recently been developed for this species \cite{Beames:JCP134:241102}. In
addition, the relatively large rotational spacing of the molecular
levels results in a molecular beam pulse with less initial population in
excited rotational states, and therefore a packet of
Stark-decelerated OH radicals with a high state purity. This facilitates
the sensitive and background free detection of scattering products and
enables the implementation of beam collimators that improve the angular
and velocity spreads of the beams at the cost of particle densities. The
disadvantage of the large rotational spacing, however, is the relatively
high energies of the energetic thresholds for inelastic scattering, and
corresponding relatively high collision energies at which scattering
resonances appear. In this respect, the inelastic scattering of OD
radicals or ND$_3$ molecules with He and Ne atoms will be interesting
candidates for studying scattering resonances as well, reducing the
energy for the lowest lying threshold to 42 cm$^{-1}$ and 14 cm$^{-1}$,
respectively. To reach a collision energy resolution $\leq$ 2 cm$^{-1}$
at these energies will relax the requirements on the velocity and
angular spreads of both beams. We will investigate resonance effects in
the rotationally inelastic scattering of OD and ammonia molecules with
He atoms in forthcoming publications.

\section{Conclusions}

We have presented detailed calculations on scattering resonances in the
rotationally inelastic scattering of OH radicals with He and Ne atoms.
For OH-He, we have developed new 3D \emph{ab initio} potential energy
surfaces, and the inelastic scattering cross sections that are derived
from these surfaces compare favorably with recent experiments. We have
identified numerous scattering resonances -- of both the shape and
Feshbach types -- in the integral cross sections. We have analyzed these
resonances using the adiabatic bender model and computed
collision lifetimes. We observe dramatic changes in the differential
cross sections at the resonances, showing in selected cases a
forward-backward peaking of the scattered flux. The analysis of
scattering resonances presented here will be indispensable in the
experimental search for such resonances in, for instance, crossed beam
scattering experiments using Stark-decelerated molecular beams. To
experimentally observe signatures of resonances in the integral cross
sections and to partially resolve individual resonances, a collision
energy resolution of $\leq$ 2 cm$^{-1}$ is required. Obtaining
energy resolutions $\leq$ 2 cm$^{-1}$ mainly requires highly collimated
molecular beams, which appears challenging. Alternatively, signatures of
scattering resonances may be found in the differential cross sections.
The selective detection of scattered molecules at backward scattering
angles may facilitate the identification of resonances if the collision
energy resolution is not sufficient to resolve them in the integral
cross sections.

\section*{ACKNOWLEDGMENTS}
KBG acknowledges support by
the European Community's Seventh Framework Program ERC-2009-AdG under
grant agreement 247142-MolChip. MHA gratefully acknowledges financial
support from the US National Science Foundation under grant No.
CHE-0848110. AvdA thanks the Alexander von Humboldt
foundation for a Humboldt Research Award. SYTM acknowledges financial support from Netherlands Organisation for
Scientific Research (NWO) via a VIDI grant. We thank our colleagues
from the Fritz-Haber-Institut in Berlin for fruitful discussions.
In particular, we thank Moritz Kirste for providing the data of Figure 5,
and Christian Schewe for help in preparing Figure 15 and discussions on
the feasibility of experimentally observing resonances using Stark-decelerated beams of OH.
We thank Gerard Meijer for carefully reading the manuscript, and for general support.

%\bibliography{heoh_resonance}

\begin{thebibliography}{54}
\bibitem[{foo({\natexlab{a}})}]{footnote1}
\bibinfo{note}{We note that parts of the resonance structure as shown in Fig. 9
  of Reference \cite{Scharfenberg2011} are unfortunately artifacts of the
  scattering calculation procedure. The conclusion in this paper that
  resonances may be observed in the experiment if collision energy
  distributions of 5 cm$^{-1}$ are achieved, is unfortunately invalid.}
\bibitem[{foo({\natexlab{b}})}]{footnote2}
\bibinfo{note}{It is noted that the resonances appear at low collision energies
  compared to the energetic threshold of the inelastic channel, rendering the
  size of the images small. A velocity map imaging detector with zoom lens
  capability, and ionization schemes that limit the ion recoil in the laser
  ionization step, will be required to measure these small images.}
\end{thebibliography}

\end{document}